\documentclass{phb-proc4m}
\usepackage{graphicx}
\usepackage{amssymb}
\usepackage{cite}

\def\abs#1{\vert #1 \vert}
\def\Mag{\langle M \rangle}
\def\Nising{\mathcal{N}^{\rm I}_{\rm conf}}
\def\Nheis{\mathcal{N}^{\rm H}_{\rm gap}}
\def\Egs{E_{\rm gs}^{\rm eff}}
\def\GS{GS}   % {ground state}
\def\VBC{VBC} % {valence-bond crystal}
\begin{document}
\begin{frontmatter}

\journal{Physica B (proceedings of SCES '04)}

%%%%%%%%%%
% Fix date of last modification
\date{14 June 2004}
%%%%%%%%%%

\title{Ground state and low-lying excitations of
the spin-1/2 $XXZ$ model on the kagom\'e lattice at magnetization 1/3}

\author[BS]{A.~Honecker\corauthref{1}},
\author[SB]{D.C.~Cabra},
\author[LP]{M.D.~Grynberg},
\author[Lyon]{P.C.W.~Holdsworth},
\author[Lyon]{P.~Pujol},
\author[MD]{J.~Richter},
\author[MD]{D.\ Schmalfu{\ss}},
\author[MD]{J.~Schulenburg}

\address[BS]{Technische Universit\"at Braunschweig,
Institut f\"ur Theoretische Physik,
Mendelssohnstrasse 3, 38106 Braunschweig, Germany}
\address[SB]{Universit\'{e} Louis Pasteur,
Laboratoire de Physique Th\'{e}orique,
3 Rue de l'Universit{\'e}, 67084 Strasbourg, C\'edex, France}
\address[LP]{Departamento de F\'{\i}sica, Universidad Nacional de La Plata,
C.C.\ 67, (1900) La Plata, Argentina}
\address[Lyon]{Laboratoire de Physique, ENS Lyon,
46 All\'ee d'Italie, 69364 Lyon C\'edex 07, France}
\address[MD]{Otto-von-Guericke-Universit\"at Magdeburg,
P.O.Box 4120, 39016 Magdeburg, Germany}

\corauth[1]{Corresponding Author:
Technische Universit\"at Braunschweig,
Institut f\"ur Theoretische Physik,
Mendelssohnstr.\ 3,
38106 Braunschweig,
Germany,
Phone: +49-531-391 5190,
Fax: +49-531-391 5833,
Email: a.honecker@tu-bs.de}

\begin{abstract}

We study the ground state and low-lying excitations of the
$S=1/2$ $XXZ$ antiferromagnet on the kagom\'e lattice
at magnetization one third of the saturation.
An exponential number of non-magnetic states is found below a magnetic gap.
The non-magnetic excitations
also have a gap above the ground state, but it is much smaller
than the magnetic gap. This ground state
corresponds to an ordered pattern with resonances in one third of
the hexagons. The spin-spin correlation function is short ranged,
but there is long-range order of valence-bond crystal type.

\end{abstract}


\end{frontmatter}

\begin{figure}[bt]
\centering
\includegraphics[width=\columnwidth]{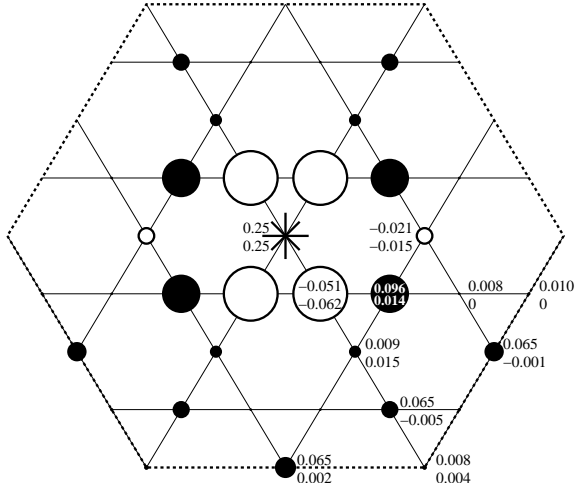}
\caption{$N=36$ kagom\'e lattice. Dashed lines on opposite sides
are identified by periodic boundary conditions.
The radius of the circle at each site $j$
shows the absolute value of the correlation
function $\abs{\langle \vec{S}_0 \cdot \vec{S}_j \rangle}$
for the $S=1/2$ Heisenberg model at $\Mag = 1/3$;
the big star denotes site $0$.
Numerical values for all inequivalent correlation functions
are shown next to one selected site $j$. The upper number corresponds
to $\langle S^z_0 S^z_j \rangle$, the lower number
to $\langle S^x_0 S^x_j \rangle = \langle S^y_0 S^y_j \rangle$.
}
\label{figcorrel}
\end{figure}

The spin $S=1/2$ Heisenberg antiferromagnet on the kagom\'e lattice has
been a focus of intense research during the past decade
since it is a hot candidate for
an exotic quantum ground state with a frustration-induced spin gap
and a continuum of singlet excitations inside this gap (see
\cite{lhuillier03,RSH04} for recent reviews).
Furthermore, a clear plateau at one third of the saturation magnetization
($\Mag = 1/3$) is
found for the $S=1/2$ $XXZ$ antiferromagnet in the presence of an external
magnetic field on the kagom\'e lattice \cite{hida,cghp,SHSRS02,km1o3}.
Here we discuss the nature of the ground state (\GS) and low-lying excitations
on this plateau.

Fig.\ \ref{figcorrel} shows the spin-spin correlation functions determined
numerically for the $S=1/2$ Heisenberg model on an $N=36$ kagom\'e lattice  
at $\Mag = 1/3$, {\it i.e.}\ in the {\GS} of the $S^z=6$
subspace. Note that the finite magnetization gives
rise to a constant contribution $1/36 \approx 0.028$ in
$\langle S^z_0 S^z_j \rangle$ and
$\abs{\langle \vec{S}_0 \cdot \vec{S}_j \rangle}$.
The nearest-neighbor correlations
$\langle S^z_0 S^z_1 \rangle \approx -0.0506$ are consistent with
an up-up-down spin arrangement around each triangle which would ideally
give rise to a value of $-1/12 \approx -0.0833$. Around a hexagon,
the sign of the correlations alternates antiferromagnetically from site
to site. In particular at larger distances $\langle S^z_0 S^z_j \rangle$
dominates over
$\langle S^x_0 S^x_j \rangle = \langle S^y_0 S^y_j \rangle$.
Despite the limited distances accessible on the $N=36$ lattice,
it is evident that the spin-spin correlations decay
rapidly and are short-ranged.

Further insight can be gained by consideration of the $XXZ$ model
\cite{km1o3}. In the Ising limit, the {\GS}s at
$\Mag =1/3$ are those states where around each triangle two spins
point up and one down. These Ising configurations can be explicitly
enumerated and the number of configurations $\Nising$
determined for a given lattice size $N$ (Table \ref{tab1} lists
a few values). The Ising configurations can be counted asymptotically
exactly for large $N$ (see \cite{km1o3,Wu} and references therein),
and one finds the growth law $\Nising \sim (1.1137\ldots)^N$
on an $N$-site kagom\'e lattice, in good agreement with the values
in Table \ref{tab1}.

For $N\le 36$,
Table \ref{tab1} also lists the number of non-magnetic states
$\Nheis$ inside the magnetic gap of the {\it Heisenberg} model,
defined by half the width of the $\Mag = 1/3$ plateau.
These numbers are slightly smaller than $\Nising$, indicating that
$\Nheis$ and $\Nising$ both grow exponentially with $N$ in a very
similar manner, and that the Ising configurations can be used to
describe the low-lying states of the Heisenberg model.

Finite values of the $XXZ$ anisotropy can be treated within
perturbation theory around the Ising limit. The induced
transitions between different Ising configurations are
described by the effective Hamiltonian \cite{MSC,km1o3}
\begin{equation}
\lambda \!\! \sum_{{\rm hexagon}\ i} \!\! \left\{
\left\vert \lower 3 mm\hbox{\includegraphics[height=0.8 true cm]{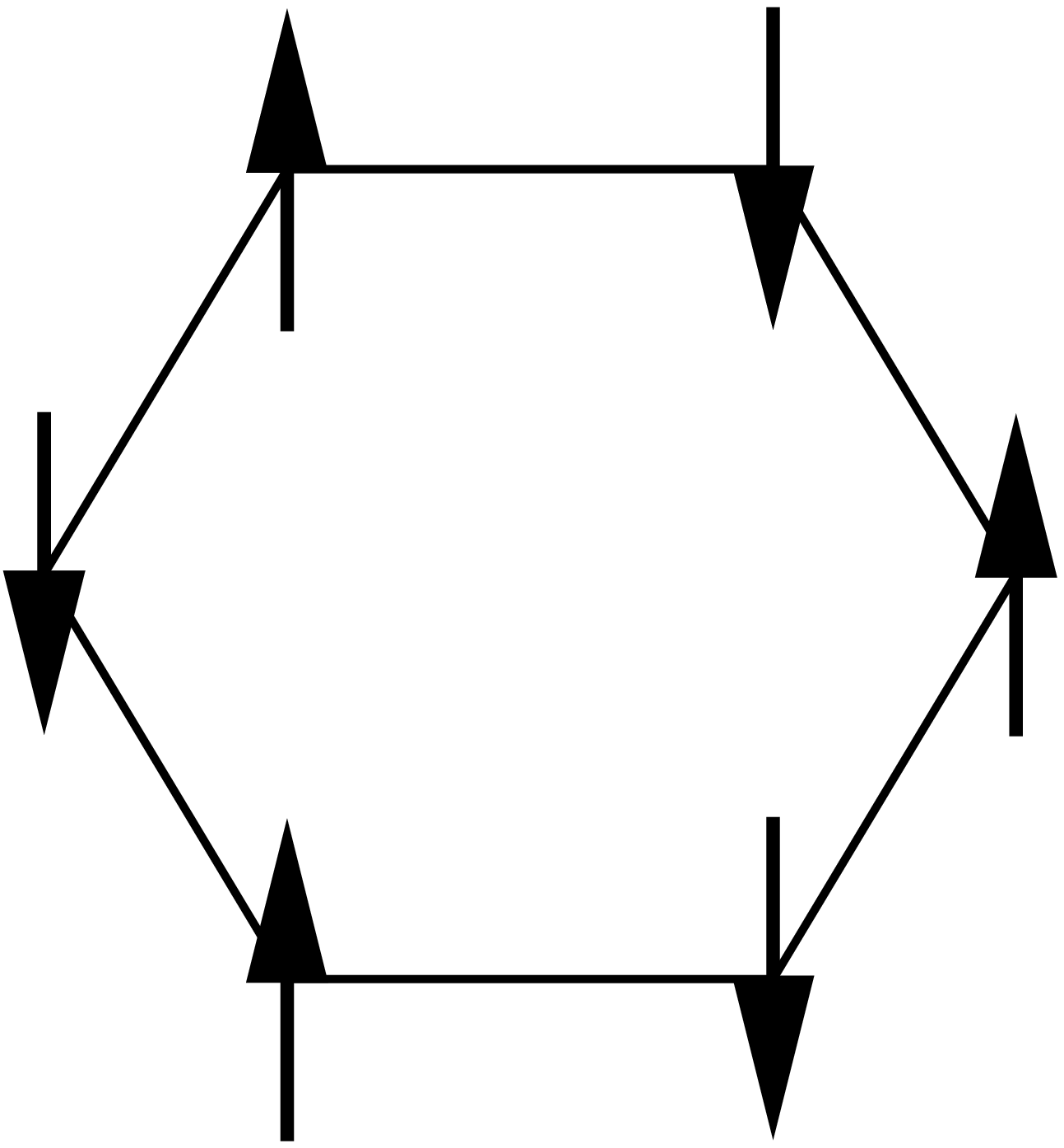}}_i \right\rangle
\left\langle \lower 3 mm\hbox{\includegraphics[height=0.8 true cm]{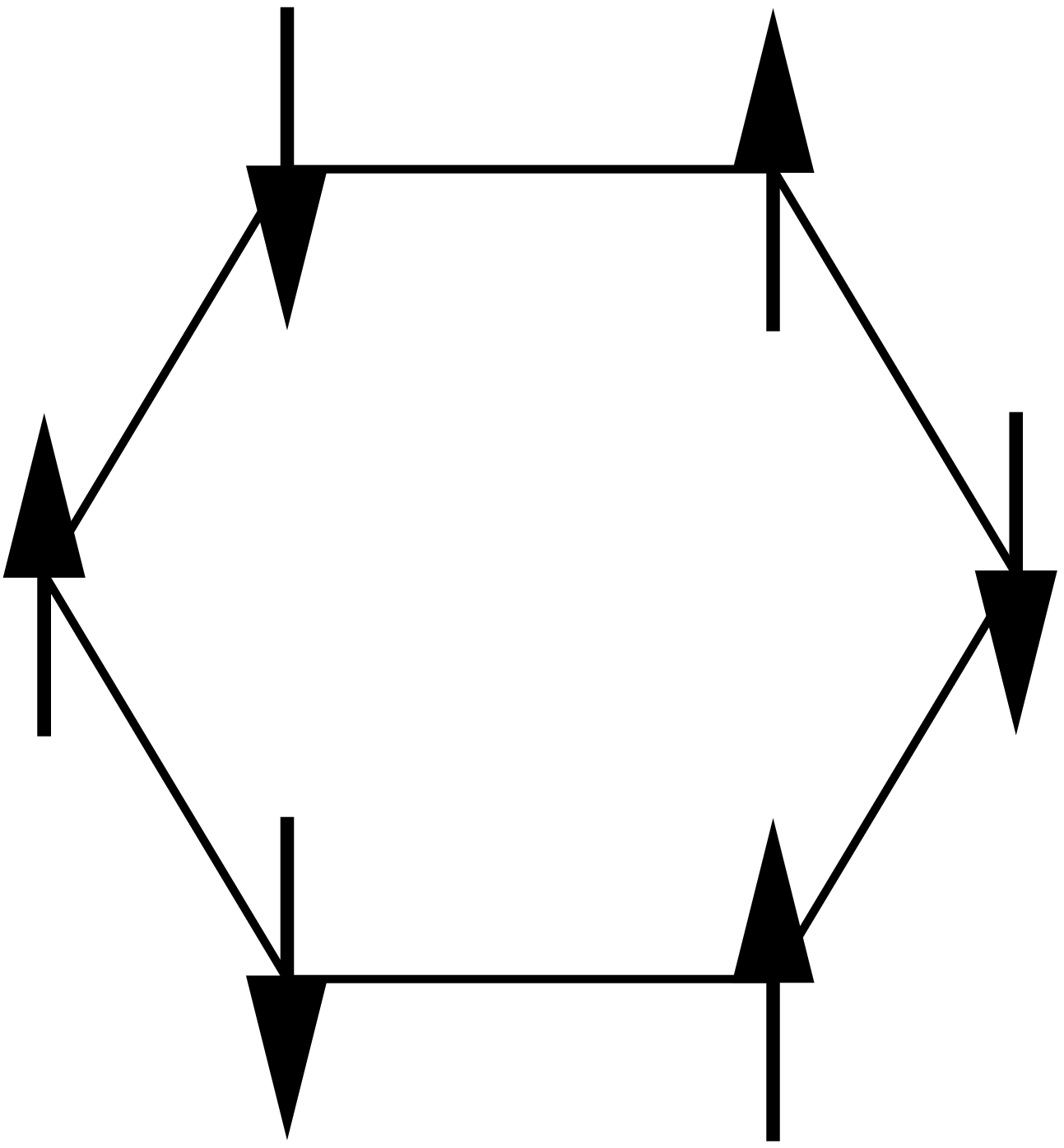}}_i \right\vert
+
\left\vert \lower 3 mm\hbox{\includegraphics[height=0.8 true cm]{honecker-204-f2inc2.eps}}_i \right\rangle
\left\langle \lower 3 mm\hbox{\includegraphics[height=0.8 true cm]{honecker-204-f2inc1.eps}}_i \right\vert
\right\} \label{Heff} 
\end{equation}
at lowest non-vanishing order and for $N\ge 36$. The sum
in (\ref{Heff}) runs over {\it all} hexagons $i$ of the kagom\'e lattice.
$\lambda > 0$ in the present context.

The effective Hamiltonian (\ref{Heff}) is equivalent to a quantum dimer model
on the hexagonal lattice whose {\GS} is
a valence-bond crystal (\VBC) \cite{MSC,MSC2001}. On the kagom\'e lattice, this
\VBC-type state can be visualized in terms of the following
three-fold degenerate variational wave function
\begin{equation}
\prod_{\sqrt{3} \times \sqrt{3} \atop {\rm hexagon}\ j} \!\!\!
  {1 \over \sqrt{2}} \left(
\left\vert
  \lower 4 mm\hbox{\includegraphics[height=1.0 true cm]{honecker-204-f2inc1.eps}}_j \right\rangle
- \left\vert
  \lower 4 mm\hbox{\includegraphics[height=1.0 true cm]{honecker-204-f2inc2.eps}}_j \right\rangle
\right)
\prod_{{\rm rest}\ k} \left\vert\
  \lower 1.5 mm\hbox{\includegraphics[height=0.4 true cm]{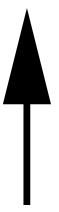}}_k \right\rangle
 \, ,
\label{varWfunc}
\end{equation}
where the first product now runs over an ordered $\sqrt{3} \times \sqrt{3}$
pattern of non-overlapping hexagons $j$ (see inset of Fig.\ 1 of \cite{km1o3}). 
{}From (\ref{varWfunc}) one obtains the variational energy
$- \lambda N/9$.
Although this is only in moderate
agreement with the numerically exact {\GS} energies
$\Egs$ of (\ref{Heff}) shown in Table \ref{tab1},
the variational wave function (\ref{varWfunc}) is still qualitatively
correct \cite{MSC,MSC2001}.

\begin{table}[bt]
\centering
\begin{tabular}{r|c|c|l} $N$ & $\Nheis$ & $\Nising$ &
$\Egs/\lambda$ \\ \hline
18  & 13  & 20 & \\
27  & 31  & 42 & \\
36  & 100 & 120                    & $-4.690415760\ldots$ \\
54  &     & 884                    & $-6.824621992\ldots$ \\
81  &     & 15 162                 & $-9.970025439\ldots$ \\
108 &     & 281 268                & $-13.28992801\ldots$ \\
144 &     & 13 219 200             & $-17.63317376\ldots$ \\
\end{tabular}
\caption{Number of non-magnetic states $\Nheis$ in the magnetic gap
of the Heisenberg model, number of {\GS} configurations $\Nising$
of the Ising model, and {\GS} energy $\Egs$
of (\ref{Heff}) for some lattice sizes $N$.}
\label{tab1}
\end{table}

The correlation functions shown in Fig.\ \ref{figcorrel} for the
{\it Heisenberg} model are
very similar to those of the {\VBC}
{\GS} of the effective Hamiltonian (\ref{Heff}).
Further evidence that the Heisenberg model and the vicinity
of the Ising limit belong to the same phase is provided by the behavior
of the overlap of the corresponding wave functions as a function of the
$XXZ$ anisotropy parameter and comparison of the spectra of the lowest
excitations obtained numerically for the $N=36$ kagom\'e lattice \cite{km1o3}.

To summarize, the $S=1/2$ Heisenberg antiferromagnet on the kagom\'e lattice
has a three-fold degenerate {\GS} of {\VBC}-type at
$\Mag = 1/3$ and a small gap to all
excitations, although there are exponentially many non-magnetic
states inside the magnetic gap.

\end{document}